 \documentclass[final,3p,times]{elsarticle-mod}
\usepackage{amssymb,amsmath,bm}
\begin{document}

\renewcommand{\labelenumi}{(\roman{enumi})}

\begin{frontmatter}

\title{CIP/multi-moment finite volume method with arbitrary order of accuracy\footnote{This paper was published in Proceedings of the {\it 12th Computational Engineering Conference of Japan Society for Computational Engineering and Science (JSCES)} Vol.12, May 22-24, 2007, Tokyo. Part of the contents were also published in a Japanese monograph {\it Computational Fluid Dynamics}, ISBN 978-4-339-04597-0, publish by Coronasha, 2009.}}

\author[titech]{Feng Xiao }\corref{cor}
\author[titech,uosaka]{Satoshi Ii}

\address[titech]{Department of energy sciences, Tokyo Institute of Technology,   
4259 Nagatsuta, Midori-ku, Yokohama, 226-8502, Japan}
\address[uosaka]{Current address: Department of Mechanical Science \& Bioengineering, Osaka University, Toyonaka, Osaka, 560-8531, Japan}

\cortext[cor]{Corresponding author: Feng Xiao, \it{xiao@es.titech.ac.jp}}

\begin{abstract}
%% Text of abstract
This paper presents a general formulation of the
CIP/multi-moment finite volume method (CIP/MM FVM) for arbitrary
order of accuracy. Reconstruction up to arbitrary order can be built
on single cell by adding extra derivative moments at the cell
boundary. The volume integrated average (VIA) is updated via a
flux-form finite volume formulation, whereas the point-based
derivative moments are computed as local derivative Riemann problems
by either direct interpolation or approximate Riemann solvers.
\end{abstract}

\begin{keyword}
High order scheme \sep finite volume method \sep multi-moment \sep fluid dynamics \sep derivative Riemann problem \sep conservation

\end{keyword}

\end{frontmatter}

%%
%% Start line numbering here if you want
%%
%\linenumbers

%% main text

\section{Introduction}
The multi-moment concept underlying the CIP method
(Cubic-Interpolated  Pseudo-particle or Constrained Interpolation
Profile)\cite{ya91} provides a general methodology to construct
numerical schemes with great flexibility. One of the major outcome
from the practice so far to implement the multi-moments in
computational fluid dynamics is that we can build high order schemes
on a relatively compact grid stencil using multi-moments, and these
moments can be carried forward in time separately by completely
different numerical approaches.

Some schemes have been developed for practical use based on VIA and
SIA (Surface-Integrated Average) \cite{xi04}\cite{xi05}
 and on VIA and PV (Point Value) \cite{ii05}\cite{ix07a}.
 The later is much more suitable for
unstructured or other complex computational grids where a point-wise
local Riemann problem can be posed at any specified point to update
the PV. It is found that increasing the number of the PVs is a
simple way to get higher order schemes. We have devised and verified
the schemes up to 4th order on 2D triangular unstructured grid for
both scalar and system conservation laws by employing both VIA and
PV moments. On the other hand, making use of the first derivative at
the cell boundary as another moment has been ever used in the
so-called CIP-CSL4(CIP-Conservative Semi-Lagrangian with 4th order
polynomial) advection scheme \cite{tny00}.

We in this paper explore further the possibility to construct
conservative CIP/multi-moment formulation of arbitrary order over
single cell using more derivative moments. The spatial
reconstruction based on multi-moments is described in section 2.
 The numerical formulation for scalar hyperbolic conservation
law is presented in section 3. The extension to Euler equations is
discussed in section 4.  Section 5 ends the paper with a few
conclusion remarks.
%%%%%%%%%%%%%%%%%%%%%%%%
% Section
%%%%%%%%%%%%%%%%%%%%%%%%
\section{The multi-moment spatial reconstruction }

The essential point in high resolution scheme is how to reconstruct
the interpolation function to find the numerical flux at the
boundary of each grid cell. Among the most widely used are, for
example, the MUSCL scheme \cite{va79}, the ENO scheme\cite{ha87} and
the WENO scheme \cite{ji96}. In all of these schemes the
interpolation is based only on the cell-averaged values of the
physical field to be reconstructed. In this section, we describe a
numerical interpolation that makes use of not only the
volume-integrated average over each mesh cell but also the
derivatives at the cell boundary. We call the present formulation
the ``multi-moment'' reconstruction to distinguish it from the
aforementioned ones which should be more properly refer to as the
``single-moment'' reconstruction.

We consider a physical field variable  $\phi(x)$ over a
one-dimensional domain divided into control volumes (mesh cells)
$[x_{i-1/2},x_{i+1/2}]$; $i=1,2,..., I$.

There is a flexibility in choosing the discretised  moments for
field variable  $\phi(x,t)$. The primary moment of the finite volume
method is the volume-integrated average (VIA) over each mesh cell
 \begin{equation}
  \overline{^V \phi}_i=\frac{1}{\Delta x_i} \int^{x_{i + \frac{1}{2}}}_{x_{i - \frac{1}{2}}} \phi(x,t) dx
 \end{equation}
where $\Delta x_i=x_{i+1/2} - x_{i-1/2}$.  The spatial derivatives
up to $K$th order at cell boundary
 \begin{eqnarray}
  \overline{^{D_x(k)} \phi}_{i + \frac{1}{2}} &=& \frac{\partial^k \phi}{\partial x^k}(x_{i + \frac{1}{2}},t)
  \equiv\partial^{(k)}_x \phi;\\ \nonumber
   &&\quad  {\rm with} \quad k=0,1,\cdots,K;
 \end{eqnarray}
are also used as the moments. Note that the  point value (PV)
$\overline{^P \phi }_{i+1/2}$ is actually equivalent to the $0$th
derivative moment $\overline{^{D_x(0)} \phi}_{i + \frac{1}{2}}$. We
will denote the $k$th derivative of $\phi$ in respect to any
variable $\beta$, $\partial^k\phi/\partial \beta^k $,  by
$\partial^{(k)}_\beta \phi$ occasionally hereafter.

Given one VIA $\overline{^V \phi}_i$ and $2(K+1)$  derivative
moments  $\overline{^{D_x(k)} \phi}_{i\pm \frac{1}{2}} $ over
$[x_{i-1/2},x_{i+1/2}]$, as well as the first-order derivative or
gradient  $d_i$ that is computed in terms of other independent
moments,  we can construct a $[2(K+1)+1]$th order cell-wise
polynomial $\Phi_i(x,t)$ with $2(K+2)$ constrained conditions as
follows,
\begin{eqnarray}
  & & \int^{x_{i + \frac{1}{2}}}_{x_{i - \frac{1}{2}}} \Phi_i(x,t) dx=\overline{^V \phi}_i, \label{constraint1} \\
  & & \partial^{(k)}_x \Phi_i (x_{i+\frac{1}{2}})=\overline{^{D_x(k)} \phi}_{i+\frac{1}{2}}, \quad k=0,1,\cdots,K; \label{constraint2} \\
  & & \partial^{(k)}_x \Phi_i (x_{i-\frac{1}{2}})=\overline{^{D_x(k)} \phi}_{i-\frac{1}{2}}, \quad k=0,1,\cdots,K;\label{constraint3} \\
  & & \partial^{(1)}_x \Phi_i (x_i)=d_i.\label{constraint4}
\end{eqnarray}

Thus, the piecewise interpolation polynomial,
\begin{equation}
 \Phi_i (x) = \sum_{k=0}^K a_k (x-x_{i-\frac{1}{2}})^k, \quad k=0,1,\cdots,K
\label{interpol}
\end{equation}
 is constructed over cell $i$. All the coefficients $a_k$ can be uniquely computed
 from (\ref{constraint1}),(\ref{constraint2}),(\ref{constraint3}) and (\ref{constraint4}).
\begin{figure}[h]
\begin{center}
\hspace{0.0cm} \includegraphics[width=7cm]{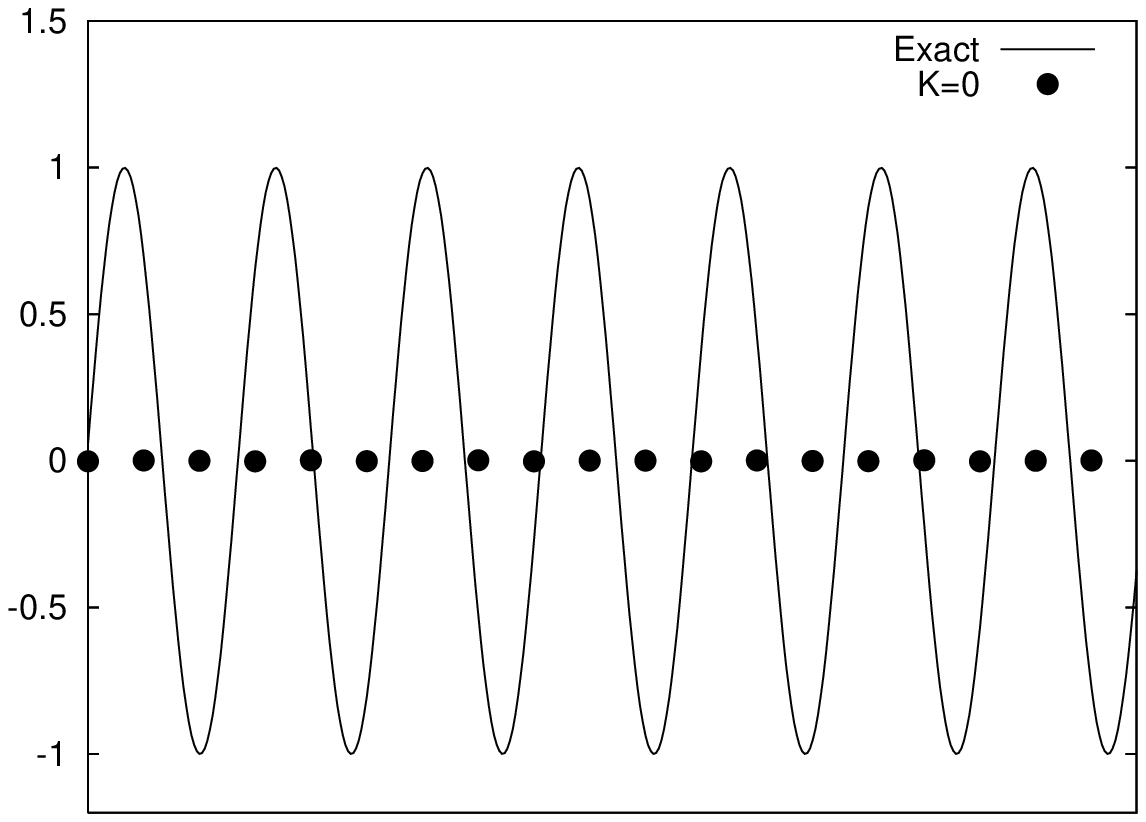} 
\hspace{0.0cm} \includegraphics[width=7cm]{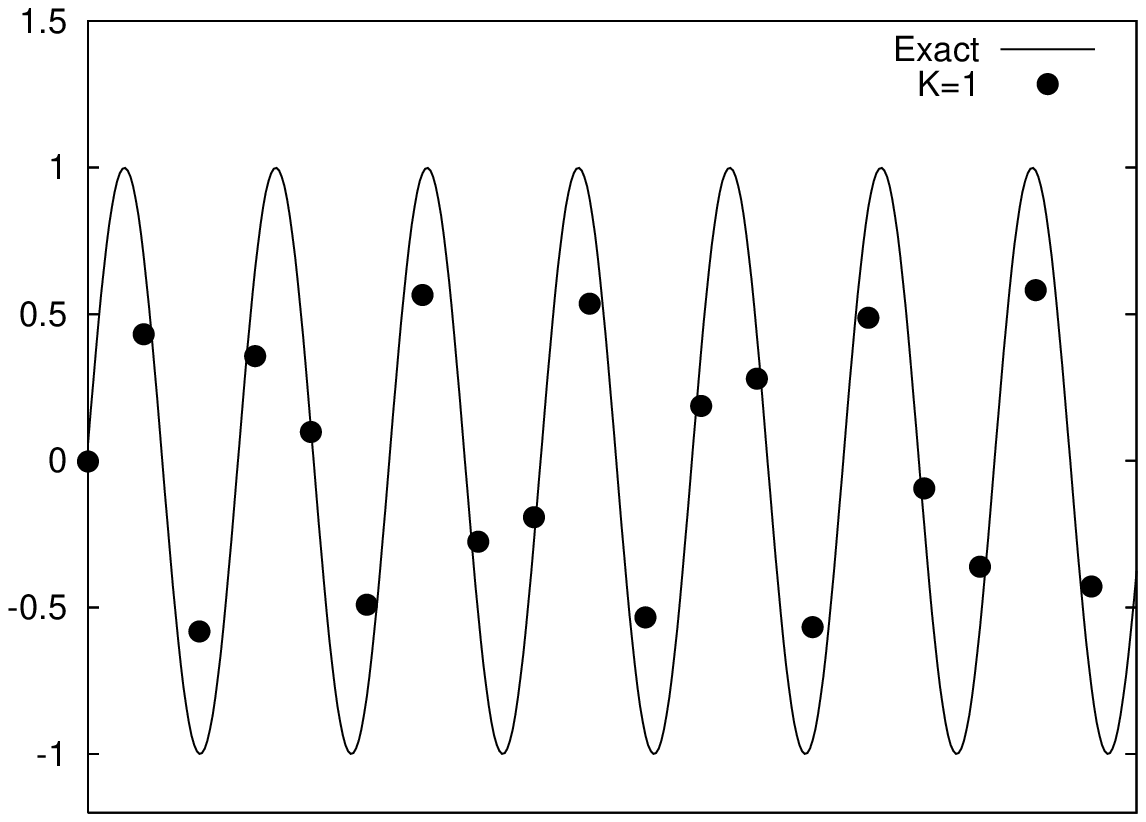} 
\hspace{0.0cm} \includegraphics[width=7cm]{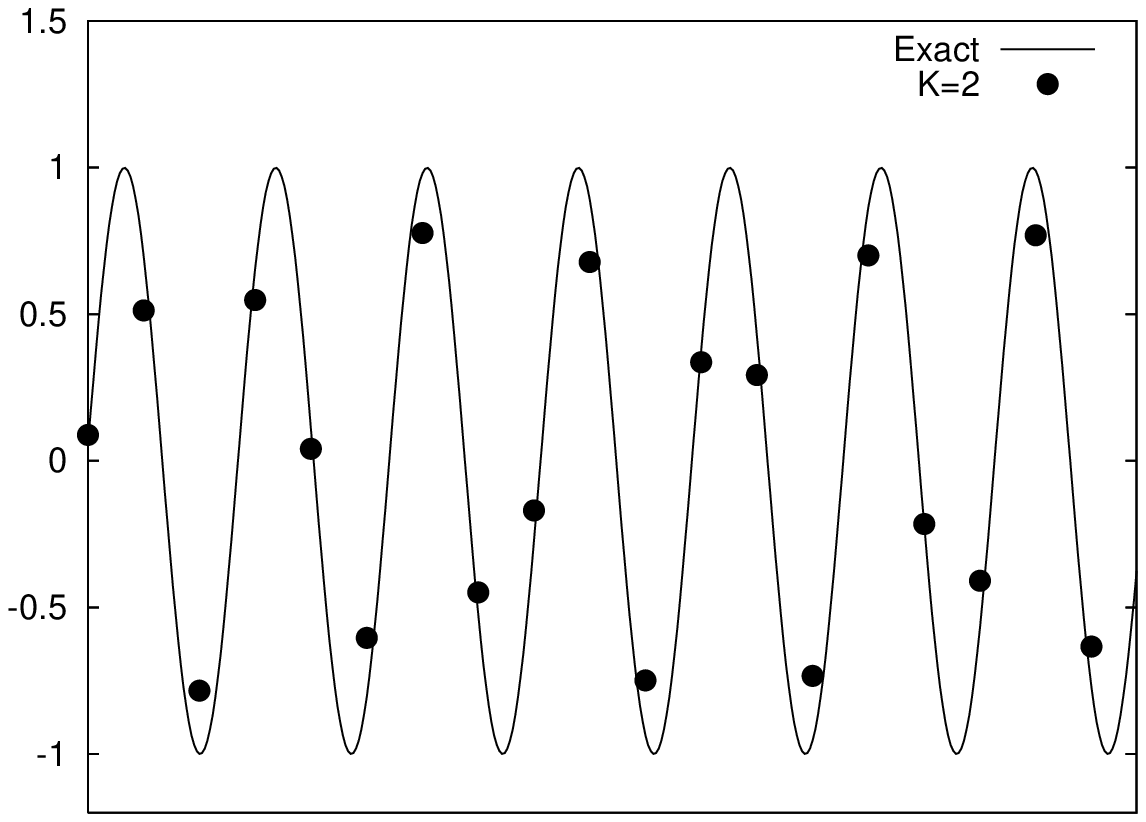} 
\hspace{0.0cm} \includegraphics[width=7cm]{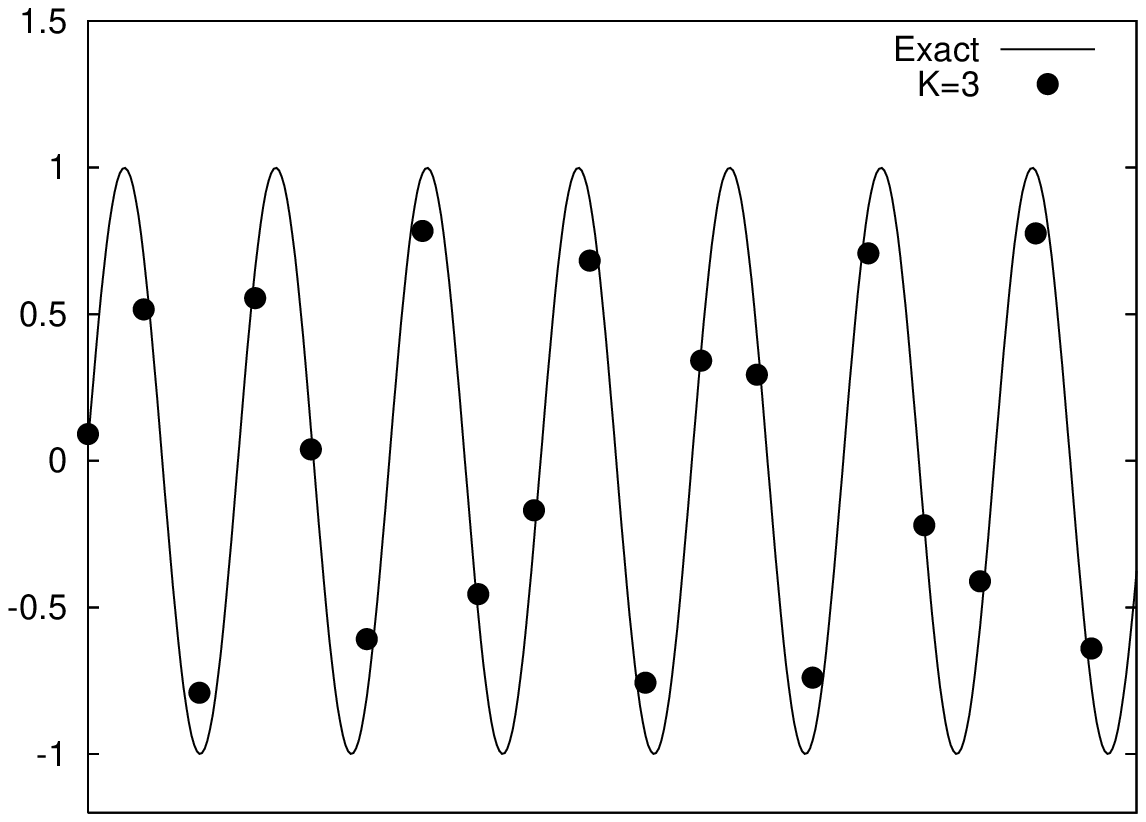} 
\caption{ Advection transport of a harmonic wave after 400 steps
(CFL=0.1) by schemes with derivative moments of 0th order
(top-left), 1th order (top-right), 2nd order (bottom-left) and 3rd
order (bottom-right). }
\end{center}
\label{fig:1}
\end{figure}
The first-order derivative or gradient of the interpolation function
$d_i$ can be approximated in terms of the known moments. For
example,  a $[2(K+1)+1]$th order polynomial is obtained if we
specify $ d_i = \partial^{(1)}_x {\tilde \Phi}_i (x_i)$ with
${\tilde \Phi}_i(x)$ is computed from constraint conditions
(\ref{constraint1}), (\ref{constraint2}) and (\ref{constraint3}).
Furthermore, a slope limiting can be imposed to $d_i$ to suppress
the numerical oscillation (see \cite{xi01} for details). We used a
single-cell minmod limiter in this paper.

\begin{figure}[h]
\begin{center}
\hspace{0.0cm} \includegraphics[width=7cm]{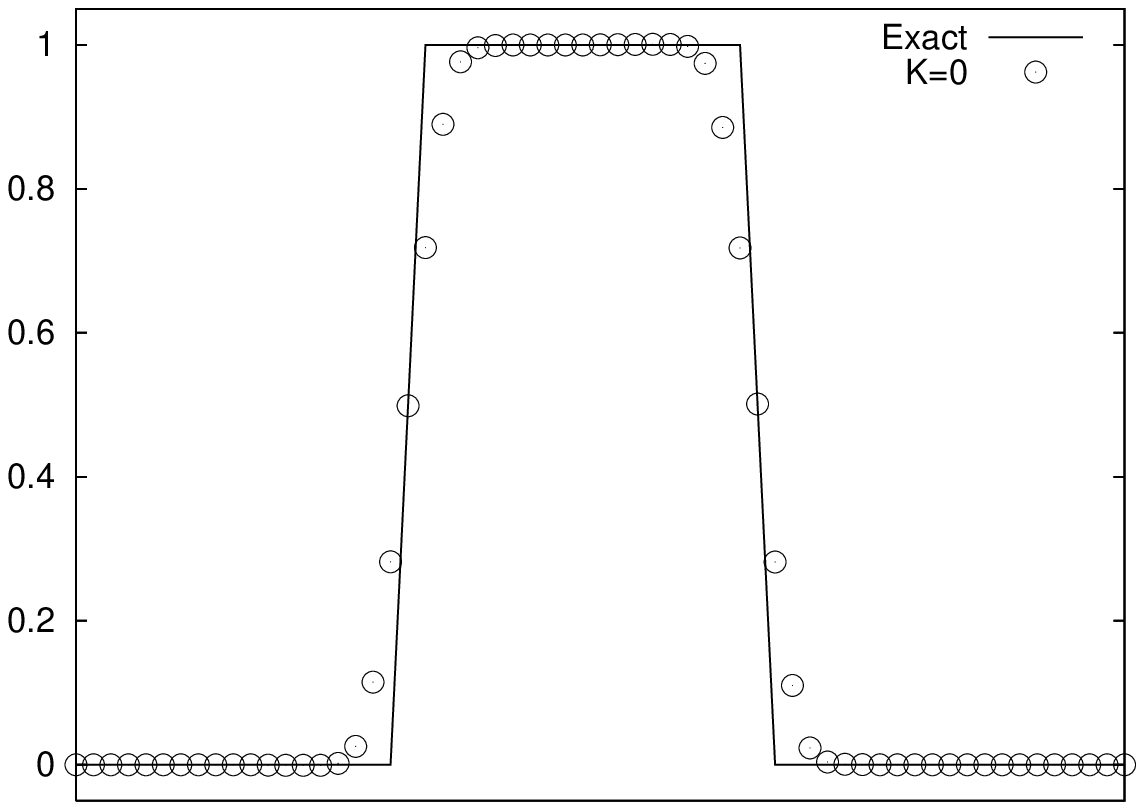} 
\hspace{0.0cm} \includegraphics[width=7cm]{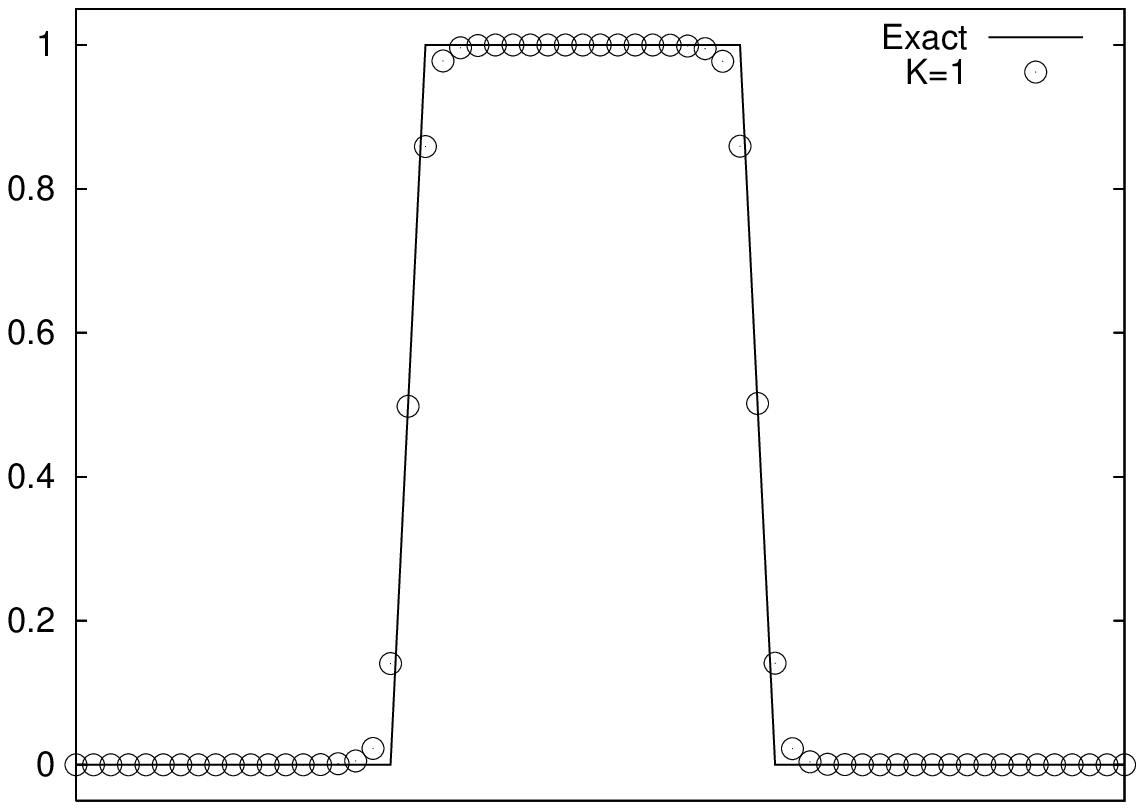} 
\hspace{0.0cm} \includegraphics[width=7cm]{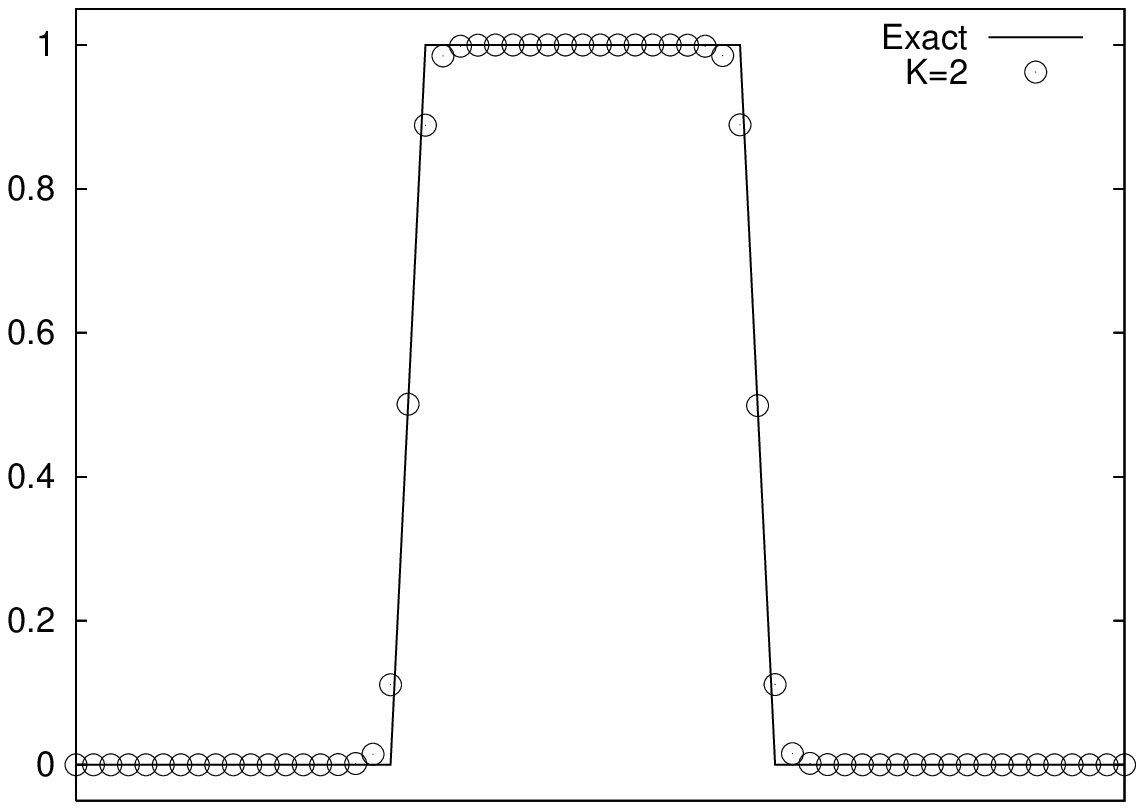} 
\hspace{0.0cm} \includegraphics[width=7cm]{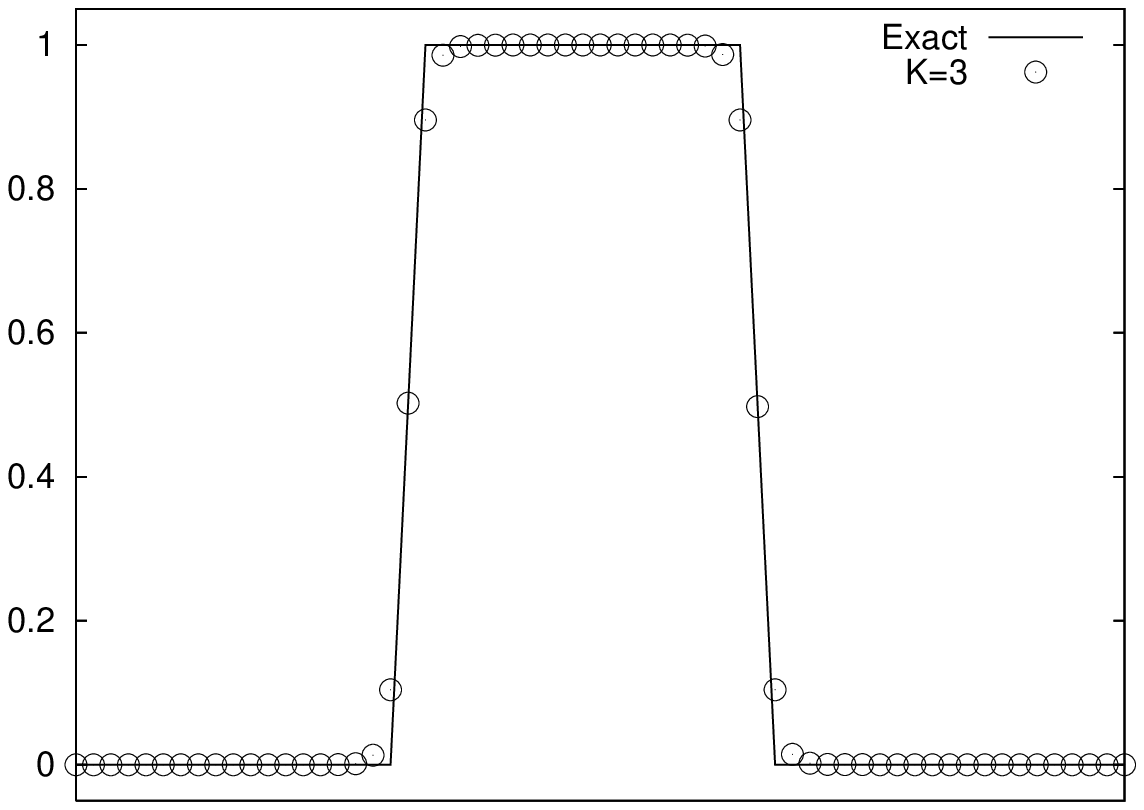} 
\caption{ Advection transport of a square pulse after 2000 steps
(CFL=0.1) by schemes with derivative moments of 0th order
(top-left), 1th order (top-right), 2nd order (bottom-left) and 3rd
order (bottom-right).}
\end{center}
 \label{fig:1}
\end{figure}
It is obvious that the reconstruction discussed above can have a
$2(K+1)$th order accuracy for smooth solutions.

\section{The scalar conservation laws}
In this section, we consider  the scalar conservative law as
follows,
\begin{equation}
 \frac{\partial \phi}{\partial t} + \frac{\partial f(\phi)}{\partial x} = 0,
\label{scalar-eq}
\end{equation}
where $\phi$ is the scalar state variable and $f(\phi)$ is the flux
function. Assuming the hyperbolicity, we have a real  characteristic
velocity, $u=\partial^{(1)}_\phi f(\phi)$.

The governing equations for the derivative moments can be directly
derived from (\ref{scalar-eq}) as,
\begin{eqnarray}
{\partial^{(k)}_t}\left (\overline{^{D_x(k)}
\phi}_{i+\frac{1}{2}}\right )=-
 {\partial^{(k+1)}_x}\left ({\hat f} (\overline{^{D_x(k)} \phi})\right )_{i+\frac{1}{2}}, \\ \nonumber
 \quad k=0,1,\cdots,K;
 \label{dv-eq1}
\end{eqnarray}
where ${\hat f}$ is the numerical flux function consistent to $f$.

It is observed from the reconstruction that the derivatives moments
up to the $K$th order $\overline{^{D_x(k)} \phi}$ and flux
$\partial^{(k)}_x {\hat f}$ are continuous.  Thus, we can update the
derivative moments $\overline{^{D_x(k)} \phi}_{i+\frac{1}{2}}$ for
$k=0,1,\cdots,K-1$ by (\ref{dv-eq1}) with the spatial derivatives of
the flux function  $\partial^{(k)}_x {\hat f}_{i+\frac{1}{2}}$
directly computed from the derivative moments that readily defined
and computed at the cell interface as
\begin{equation}
\partial^{(k)}_x {\hat f}_{i+\frac{1}{2}}= \partial^{(k)}_x { f} {\big
(} (\overline{^{D_x(0)} \phi}_{i+\frac{1}{2}},\overline{^{D_x(1)}
\phi}_{i+\frac{1}{2}},\cdots,\overline{^{D_x(k)}
\phi}_{i+\frac{1}{2}} {\big )}. \label{contin-flux}
\end{equation}

When one advances the highest order derivative moment
$\overline{^{D_x(K)} \phi}$ , $\partial^{(K+1)} {\hat f}/\partial
x^{K+1}$ is required, which, however, might not be continuous at the
cell boundaries.  We make use the the simple Lax-Friedrichs
splitting in terms of the spatial derivatives of the flux function
and the state variable as
\begin{eqnarray}
&& \partial^{(K+1)}_x {\hat f}_{i+\frac{1}{2}}= \\\nonumber
&&\frac{1}{2}\left( \partial^{(K+1)}_x {\hat
f}(\phi^L)_{i+\frac{1}{2}}+
\partial^{(K+1)}_x {\hat f}(\phi^R)_{i+\frac{1}{2}}\right ) \\\nonumber
&&-\frac{\alpha_{i+\frac{1}{2}}}{2} \left( \partial^{(K+1)}_x
{\phi}^L_{i+\frac{1}{2}}-\partial^{(K+1)}_x
{\phi}^R_{i+\frac{1}{2}}\right ), \label{split}
\end{eqnarray}
where $\alpha_{i+\frac{1}{2}}$ is the largest value of the
characteristic speed in the related region.

The state variables $\phi^L_{i+\frac{1}{2}}$ and $\hat
\phi^R_{i+\frac{1}{2}}$ are  computed from the multi-moment
reconstructions (\ref{interpol}) separately built for cells
$[x_{i-1/2},x_{i+1/2}]$ and  $[x_{i+1/2},x_{i+3/2}]$, i.e.
$$  {\phi}^L_{i+\frac{1}{2}}=\Phi_i(x_{i+\frac{1}{2}}) \quad  {\rm  and} \quad  {\phi}^R_{i+\frac{1}{2}}=\Phi_{i+1}(x_{i+\frac{1}{2}}).$$
The corresponding $(K+1)$ derivatives are
\begin{eqnarray}
&& \partial^{(K+1)}_x{\phi}^L_{i+\frac{1}{2}}=\partial^{(K+1)}_x
\Phi_i(x_{i+\frac{1}{2}})  \quad {\rm  and} \\ \label{left_k1}
&& \partial^{(K+1)}_x{\phi}^R_{i+\frac{1}{2}}=\partial^{(K+1)}_x
\Phi_{i+1}(x_{i+\frac{1}{2}}). \label{right_k1}
\end{eqnarray}

It should be noted that we have used an assumption similar to
\cite{toro02}\cite{toro05a}\cite{toro05b} in getting a homogeneous
and linearized Riemann problem for spatial derivatives for the state
variable.

In order to update the VIA moment, we integrate (\ref{scalar-eq})
over $[x_{i-1/2},x_{i+1/2}]$, yielding the following conservative
formulation,
\begin{equation}
 \frac{\partial \overline{^V \phi}_i}{\partial t} = - \frac{1}{\Delta x_i}
 \big({\mathcal{F}}(\overline{^{D_x(k)} \phi}_{i+\frac{1}{2}})-{\mathcal{F}}(\overline{^{D_x(k)} \phi}_{i+\frac{1}{2}}) \big),
\label{eq:fvm}
\end{equation}
where $\mathcal{F}$ denotes the numerical flux at cell boundary, and
is computed directly from the derivative moments readily updated at
the cell boundaries.

The semi-discretized time evolution equations (9) and (\ref{eq:fvm})
are predicted in time by a TVD\cite{sh88} or a 4th order Runge-Kutta
method. At every substep, we first update the derivative moments by
(9), and then use these updated moments to evaluate the flux
function in (\ref{eq:fvm}).

We computed an advected harmonic wave with a wavelength of
$20/7\Delta x$. Fig.1 shows the results for reconstructions using
derivatives moments up to different orders. It is found that even a
short wave can be adequately resolve if higher order derivative
moments are used.

Fig.2 reveals the effect of limiting. The numerical oscillation
associating discontinuities are eliminated by the slope switching
that is also constructed within a single mesh element.

\section{The Euler conservation laws}
In this section, the numerical formulation presented above is
implemented to the inviscid Euler conservation laws.

The conservative form of the one-dimensional Euler equations is
written  as follows,
\begin{equation}
 \frac{\partial \mathbf{U}} {\partial t} + \frac{\partial \mathbf{F}} {\partial x} =0, \quad
 \mathbf{U} =
  \left( \begin{array}{c}    \rho \\    \rho u \\    e \end{array}\right), \quad
 \mathbf{F} =
  \left (\begin{array}{c}
   \rho u  \\
   \rho u^2 + p \\
   u (e+p)
 \end{array} \right ),
\label{eq:Euler_1D}
\end{equation}
where $\mathbf{U}$ is the vector of conservative variables and
$\mathbf{F}$ is the vector of inviscid fluxes. Denoted by $\rho$ is
the density, $u$ the velocity, $e$ the total energy and $p$ the
pressure that is obtained by the equation of state for the perfect
gas $p=(e-\rho u^2/2)(\gamma-1)$. The ratio of the specific heats
$\gamma$ is specified as 1.4 in this paper.

The volume-integrated average (VIA) moment over mesh cell $i$,
 \begin{equation}
  \overline{^V \mathbf{U}}_i=\frac{1}{\Delta x_i}
  \int^{x_{i + \frac{1}{2}}}_{x_{i - \frac{1}{2}}} \mathbf{U}(x,t)
  dx,
 \end{equation}
and the derivative moments up to $K$th order at cell boundary,
 \begin{equation}
  \overline{^{D_x(k)} \mathbf{U}}_{i + \frac{1}{2}} = \frac{\partial^k \mathbf{U}}{\partial x^k}(x_{i + \frac{1}{2}},t)  \quad  {\rm with} \quad k=0,1,\cdots,K;
 \end{equation}
are treated as the model variables.

\begin{figure}[h]
\begin{center}
\hspace{0.0cm} \includegraphics[width=10cm]{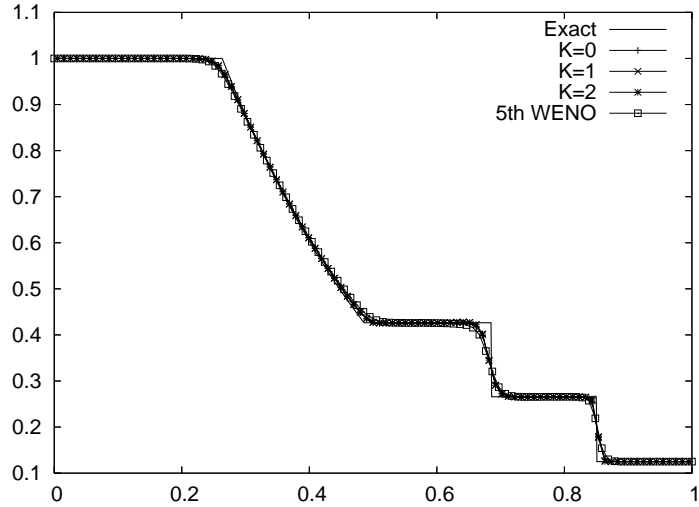} 
\caption{Numerical results of Sod's problem at $t=0.2$ separately
computed by the multi-moment reconstructions with $K=0,1,2$ and the
 5th-order WENO scheme are plotted. }
\end{center}
\end{figure}

Analogous to the scalar case, the governing equations for the
derivative moments are derived from (\ref{eq:Euler_1D}) as,
\begin{eqnarray}
&&{\partial^{(k)} t}\left (\overline{^{D_x(k)}
\mathbf{U}}_{i+\frac{1}{2}}\right )= \\ \nonumber &&
-{\partial^{(k+1)}_x}\left (\hat{ \mathbf{F}} (\overline{^{D_x(k)},
\mathbf{U}})\right )_{i+\frac{1}{2}}, \quad k=0,1,\cdots,K;
 \label{dv-eq2}
\end{eqnarray}
where $\hat{ \mathbf{F}}$ is the vector of the numerical flux
function consistent to $\mathbf{F}$.

Given $\overline{^{D_x(k)} \mathbf{U}}_{i+\frac{1}{2}}$  continuous
for $k=0,1,\cdots,K$,  we can update the derivative moments
$\overline{^{D_x(k)}\mathbf{U}}_{i+\frac{1}{2}}$ for
$k=0,1,\cdots,K-1$ by (\ref{dv-eq2}) with the spatial derivatives of
the flux function ${\partial^{(k)}_x} \hat {\mathbf{F}}$
 evaluated directly from
\begin{eqnarray}
&&\left(\partial^{(k)}_x \hat {\mathbf{F}}\right)_{i+\frac{1}{2}}= \\ \nonumber
&& \partial^{(k)}_x \hat {\mathbf{F}} {\big (}
\overline{^{D_x(0)} \mathbf{U}}_{i+\frac{1}{2}},\overline{^{D_x(1)}
\mathbf{U}}_{i+\frac{1}{2}},\cdots,\overline{^{D_x(k)}
\mathbf{U}}_{i+\frac{1}{2}} {\big )}. \label{contin-flux2}
\end{eqnarray}

Similar to the scalar case, $\partial^{(K+1)}_x \hat {\mathbf{F}}$
might be not continuous at the cell boundaries. So, the highest
order derivative moment $\overline{^{D_x(K)} \mathbf{U}}$ has to be
solved from a Riemann problem in terms of the spatial derivative. To
this end, we use the linearization assumption in
\cite{toro02}\cite{toro05a} and write the $(K+1)$th spatial
derivative of the flux function as
\begin{equation}
\partial^{(k)}_x \mathbf{F}=\mathbf{A}\partial^{(k)}_x \mathbf{U}, \label{linear-f}
\end{equation}
where $\mathbf{A}$ is the Jacobian matrix.

The conventional flux splitting algorithms can be adopted here in
terms of the spatial derivative quantities.

The $(K+1)$th derivatives of the state variables and the flux
function
 at the cell interface are  computed from the multi-moment
reconstructions separately built over two neighboring cells as (12)
and (13) component-wisely in terms of the the state variables or
characteristic variables.

\begin{figure}[!htbp]
\begin{center}
\hspace{0.0cm} \includegraphics[width=8cm]{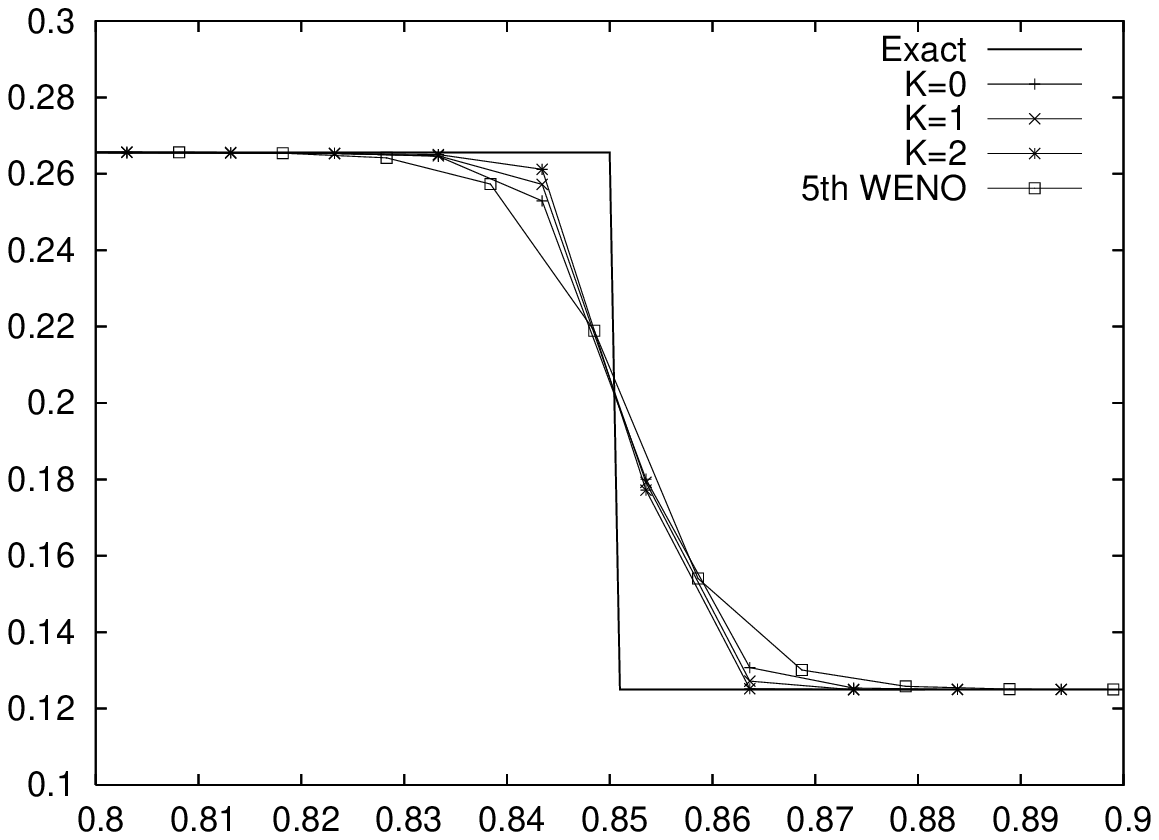} 
\hspace{0.0cm} \includegraphics[width=8cm]{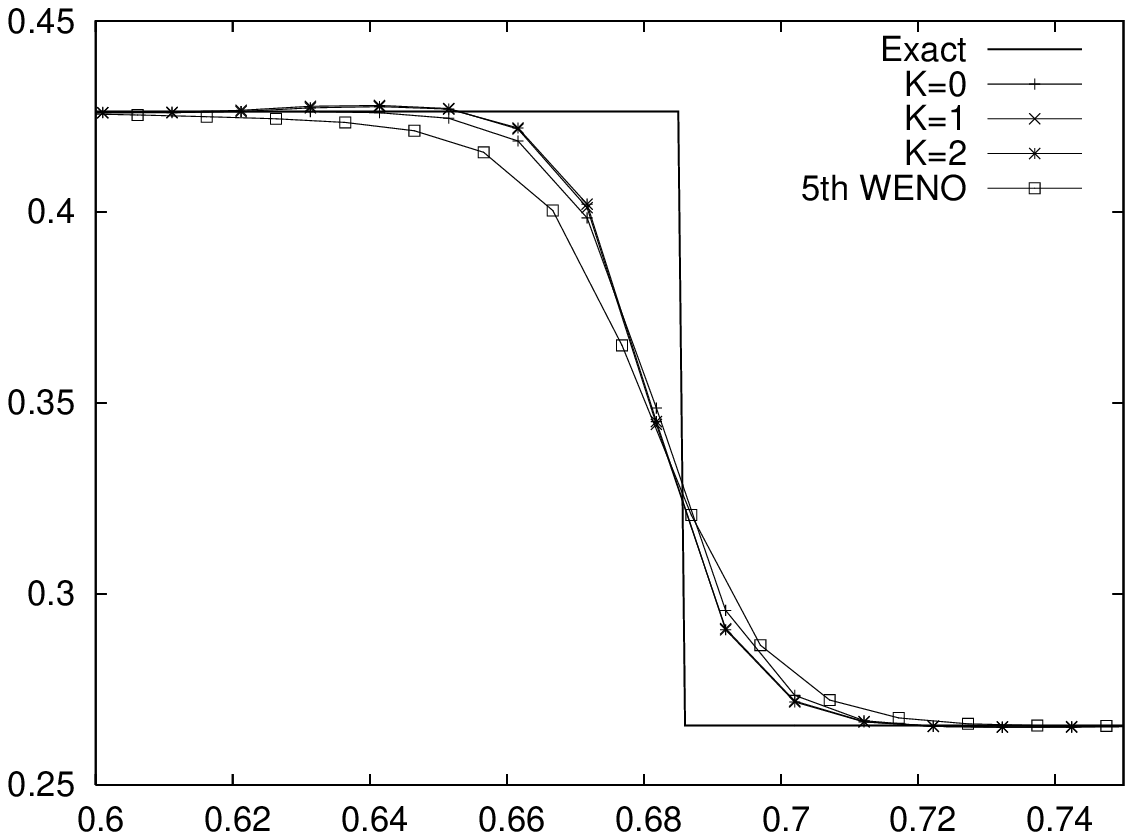} 
\caption{The close-up views for shock (left) and contact
discontinuity(right). }
\end{center}
\end{figure}
The VIAs of the conservative variables $\mathbf{U}$ on cell $i$ are
updated by integrating (\ref{eq:Euler_1D}) over
$[x_{i-1/2},x_{i+1/2}]$, which results in a finite volume
formulation,
\begin{equation}
\frac{\partial \overline{^V \mathbf{U}}}{\partial t}_i =  -
\frac{1}{\Delta x_i} \Big(
-{\hat {\mathbf{F}}}(\overline{^{D_x(k)} \mathbf{U}}_{i+\frac{1}{2}})
 -{\hat {\mathbf{F}}}(\overline{^{D_x(k)} \mathbf{U}}_{i+\frac{1}{2}}) \Big).
\end{equation}
Given the derivative moments $\overline{^{D_x(k)} \mathbf{U}}_{i+\frac{1}{2}}$ at cell boundaries, the numerical
fluxes ${\hat {\mathbf{F}}}(\overline{^{D_x(k)} \mathbf{U}}_{i+\frac{1}{2}})$ in the above equation are directly found.

Again, the Runge-Kutta  method is used for time integration for all moments.

A 1D shock tube test\cite{so78} was computed to verify the present
method for Euler conservation laws. We include the numerical result
of the 5th-order WENO scheme\cite{ji96} as well for comparison.
Shown in Fig.3, the numerical results of the present scheme with
different orders are quite competitive. The close-up plots for shock
and contact discontinuity are given in Fig.4. Both linear and
non-linear discontinuities are well resolved with correct locations.
It is observed that better resolution can be obtained by simply
increasing the order of the derivative moments.

\section{Concluding remarks}

A formulation that uses high order derivative moments has been
suggested and tested.
 Given all the derivative moments that are continuous at cell boundaries and updated separately, the
resulting numerical formulation is still single-cell based and quite
computationally efficient. In case that the derivative moments are
defined and continuous at the cell boundary, the numerical fluxes
can be computed directly as in the IDO scheme \cite{aoki97}, while
for the spatial derivative higher than the continuous one, we
simplify and cast it into a linearized derivative Riemann
problem\cite{toro02}. The present formulation is substantially
different from the ADER method\cite{toro02}\cite{toro05a}
\cite{toro05b} where all the derivatives are discontinuous at cell
boundaries, thus is more efficient.

Our numerical results show that the resolution of the scheme can be improved by simply increasing the order of the
derivative moments involved. With the simple slope limiting\cite{xi01}, the numerical oscillation around the
large gradient can be effectively suppressed.

Although the multi-dimension implementation remains an open problem to be further explored, one can expect the present scheme as an accurate
efficient solver for 1D conservation laws.


\begin{thebibliography}{00}
\bibitem{aoki97} T. Aoki, Comput. Phys. Commun. \textbf{102} (1997) 132.

\bibitem{ha87} A. Harten, B. Engquist, S. Osher and S. Chakravarthy,
                        J. Comput. Phys. \textbf{71} (1987) 231.
\bibitem{ii05} S. Ii, M. Shimuta and F. Xiao,
            Comput. Phys. Comm. \textbf{173} (2005) 17.
\bibitem{ix07a} S. Ii and F. Xiao,
             J. Comput. Phys. \textbf{222} (2007) 849.
\bibitem{ji96} G. Jiang and C.W. Shu,
                        J. Comput. Phys. \textbf{126} (1996) 202.
\bibitem{sh88} C.W. Shu,
            SIAM J. Sci. Stat. Comput. \textbf{9} (1988) 1073.
\bibitem{so78} G. Sod,
            J. Comput. Phys. \textbf{27} (1978) 1.
\bibitem{tny00} R. Tanaka, T. Nakamura and T. Yabe,
                        Comput. Phys. Commun. \textbf{126} (2000) 232.
\bibitem{toro02} V.A.Titarev and E.F.Toro, J. Sci. Comput. \textbf{17} (2002) 609.
\bibitem{toro05a} V.A.Titarev and E.F.Toro, J. Comput. Phys. \textbf{204} (2005) 715.
\bibitem{toro05b} E.F.Toro and V.A.Titarev, J. Comput. Phys. \textbf{202} (2005) 196.
\bibitem{va79} B. van Leer,    J. Comput. Phys. \textbf{32} (1979) 101.
\bibitem{xi04} F. Xiao,   J. Comput. Phys. \textbf{195} (2004) 629.
\bibitem{xi05} F. Xiao, R. Akoh and S. Ii,   J. Comput. Phys. \textbf{213} (2006) 31.
\bibitem{xi01} F. Xiao and T. Yabe,    J. Comput. Phys. \textbf{170} (2001) 498.
\bibitem{ya91} T. Yabe and T. Aoki,   Comput. Phys. Commun. \textbf{66} (1991) 219.
\end{thebibliography}
\end{document}